# Origins and evolutionary genomics of the novel 2013 avian-origin H7N9 influenza A virus in China: Early findings

Jiankui He*, Luwen Ning, Yin Tong

Department of Biology, South University of Science and Technology of China, Shenzhen, China, 518055

*Corresponding email: he.jk@sustc.edu.cn

Abstract

In March and early April 2013, a new Avian-Origin Influenza A (H7N9) Virus (A-OIV) emerged in the eastern China. This virus has caused global concern as a potential pandemic threat. Here we use evolutionary analysis to reconstruct the origins and early development of the A-OIV viruses. We found that A-OIV was derived from a reassortment of three avian flu virus strains, and substantial mutations have been detected. Our results highlight the need for systematic surveillance of influenza in avian, and provide evidence that the mixing of new genetic elements in avian can result in the emergence of viruses with pandemic potential in humans.

On April 1, 2013, the World Health Organization (WHO) first reported 3 human infections with the A-OIV in China. As of April 9, 2013, 24 laboratory confirmed infection cases have been reported in the eastern China. Most reported cases have severe respiratory illness and seven died since March[1]. No person-to-person transmission of the A-OIV has been found at this time, and the reported cases are not linked to each other. However, there are uncertainties about all aspects of the novel virus, including the transmissibility, virulence and origins, therefore, results in uncertainty in determining the pandemic potential of the virus, and to what extent pubic health organization should react, such as whether recommendations to stay at home or close school are need. Here we report our preliminary findings of the origins and genomic evolution of this virus. This study could aid policy decisions and developing proper treatment.

We achieved 6 newly sequenced 2013 A-OIV genomes which have been deposited to the database GISAID (www.gisaid.org). Among them, 4 isolates are A-OIV in human and 2 isolates are A-OIV in avian. Using comprehensive phylogenetic analyses [2-4], we have estimated a reconstruction of the complex reassortment history of the novel virus, summarized in Fig. 1. The A-OIV virus contains eight genes. The haemagglutinin gene (HA) comes from H7 family virus, the neuraminidase gene (NA) comes from N9 family virus, and the six internal genes (PA, NS, PB2, PB1, NP, M) come from H9N2. The HA gene segment of A-OIV is in the H7 family avian influenza lineage (Fig. 2). The HA gene segment most closely related to the A-OIV is from a H7N3 strain (A/duck/Zhejiang/11/2011). The NA gene segment of A-OIV is in the clade of N9 family avian influenza. The two strains that are closely related to NA gene segment A-OIV are A/mallard/Czech Republic/13438-29K/2010(H11N9) and a H7N9 strain found in Korea wild duck. We found that the six internal gene segments are highly conserved with the H9N2 (Figs. S6-S14). Phylogenetic analyses indicate that the six internal gene segments of A-OIV are derived from H9N2 Avian influenza with sequence identity >98%, in particular, the A/chicken/Zhejiang/Q1D4/2011(H9N2) strain, which was found in Zhejiang province of China, the same location as A-OIV infection cases in human. The H9N2 flu strains have been circulating in the East Asia for many years. We also observed that the sequence of HA and NA protein of A-OIV in human are nearly identical to that of A-OIV in avian, which suggests that the same A-OIV may have the ability to infect both human and avian. Classical H7N9 viruses have been reported as early as 1999. We found that A-OIV and classic H7N9 share less similarity compared to other strains such as H7N3 and H7N1. Therefore, the A-OIV is not likely mutated from the classical H7N9. Together, our results show that the A-OIV is a new virus that stems from a reassortment of three avian influenza virus strains and the reassortment events likely happened in the eastern China.

Although the A-OIV has been reported in just one week, we already observed mutations in their genome. Within the 6 newly sequenced genomes, the HA protein of A/Shanghai/1/2013 strains have 8 amino acid differences compared to 5 other strains. This indicates that the A-OIV has significant sequence diversity, which increases our concern that genetic mutation may lead to the change of transmissibility and virulence.

There has been persisting concern that avian flu may become the next pandemic among international community. H5N1 is one such avian flu that has been extensively studied and monitored for many years. Most of the A-OIV infection cases experienced close contact with poultries before infection, and the cases appeared geographically widely in three provinces of the eastern China at the same time. The virus seems not to cause diseases in poultries and could be spreading out in poultry undetected — and thus could create a reservoir of infection that would lead to frequent sporadic human infections without warning. Yet despite widespread influenza surveillance in humans, the lack of systematic avian surveillance allowed for the undetected persistence and evolution of this potentially pandemic strain for many years. Our results highlight the need for systematic surveillance of influenza in avian, and provide evidence that the mixing of new genetic elements in avian can result in the emergence of viruses with pandemic potential in humans.

## Methods

The genome sequences of newly sequenced 6 strains of 2013 A-OIV in China were achieved in GISAID. Other avian influenza genomes of any subtype and sampling year were downloaded from the NCBI Influenza Virus Resource on April 6, 2013. The protein sequences were aligned using ClustalW with default parameters. Phylogenetic trees were inferred using the Neighbor-Joining method, plotted in TreeGraph software.

## Acknowledgements

We thank Hangzhou CDC and WHO Chinese National Influenza Center for providing sequence data. We thank Yiqi Jiang, Jingwei Qi and Yujun Zhao for data analysis. This study is supported by National Science Foundation of China (grant 31200688).

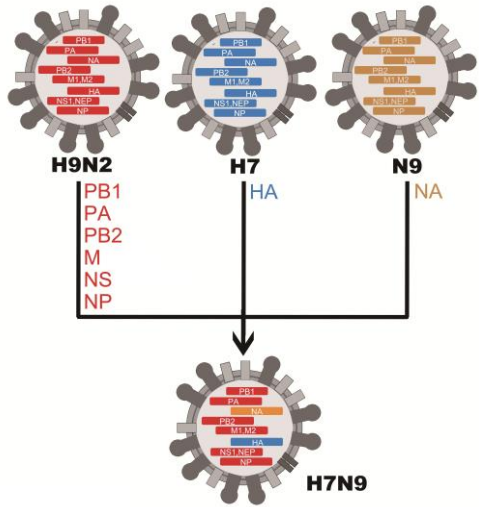

Figure 1: Reconstruction of the reassortment events leading up to the emergence of A-OIV. PB2: Polymerase basic 2; PB1: Polymerase basic 1; PA: Olymerase acidic; HA: Hemagglutinin; NP: Nucleoprotein; NA: Neuraminidase; M: Matrix gene; NS: Nonstructural gene. Color of gene segment in circle indicates host.

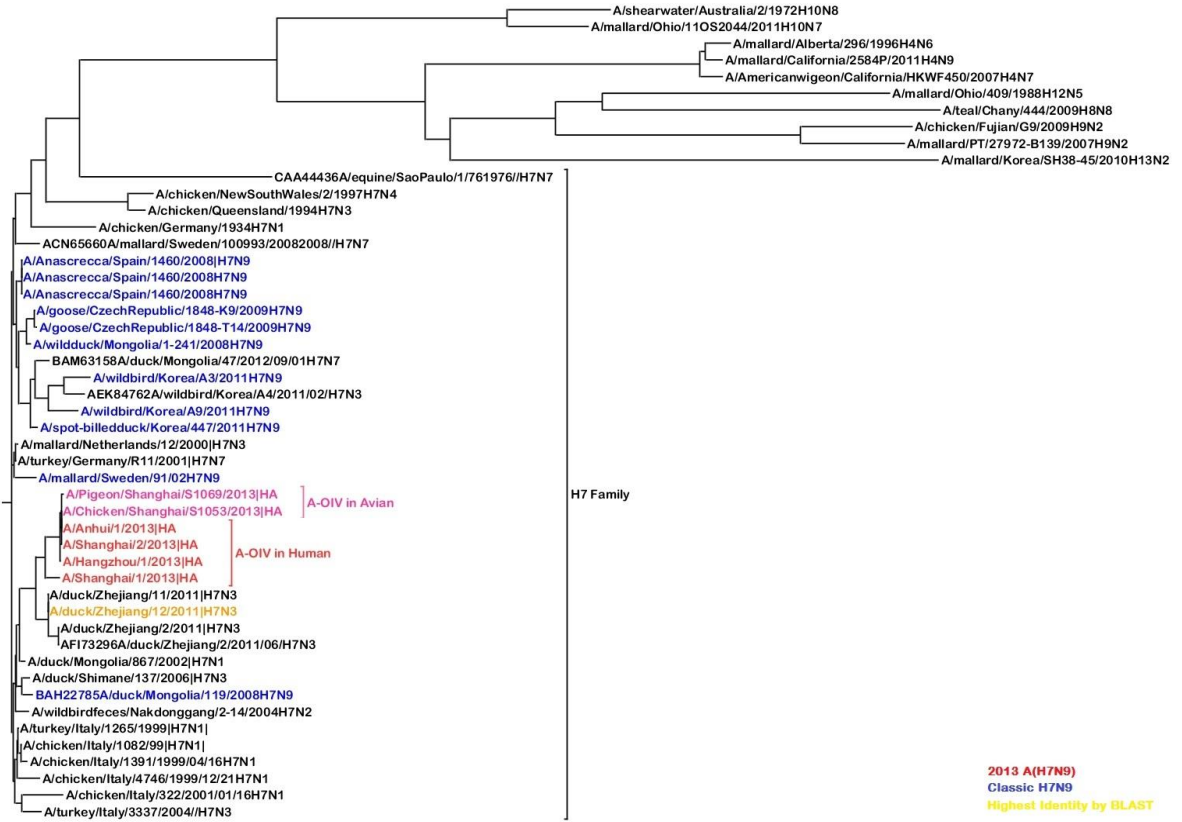

(a)

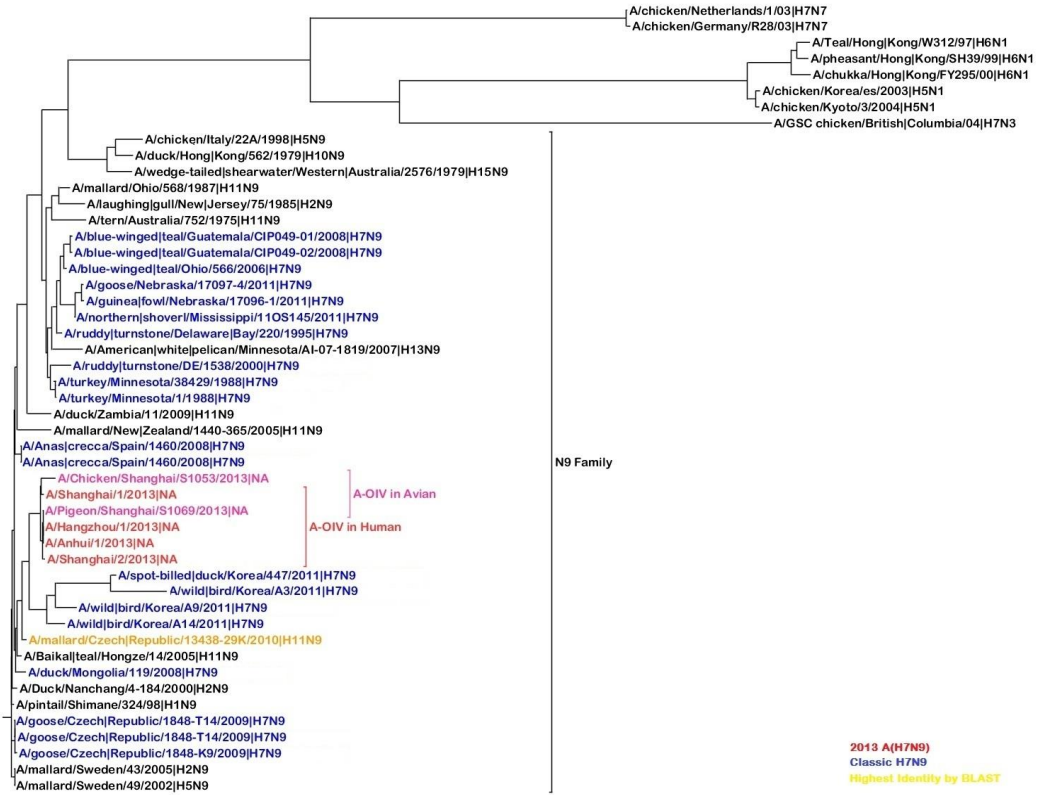

(b)

Figure 2: The phylogenetic tree for protein sequences of the (a) HA and (b) NA protein of selected influenza viruses. The selected viruses were chosen to be representative from among all available relevant sequences: A-OIV; sequences that had both high and low divergence to avoid biasing the distribution of branch lengths; strains that were representative of the major gene lineages from different subtypes; and the nearest BLAST relative to include the most closely related influenza virus in avian. The robustness of individual nodes of the tree was assessed using a bootstrap resampling analysis (1000 replicates). Comprehensive phylogenetic tree analyses of a larger number of representative every subtype, and of all avian gene segments, are shown in Figs. S4-S15.

# Origins and evolutionary genomics of the novel 2013 avian-origin H7N9 influenza A virus in China: Early analyses

## Supplemental Methods

**Section 1: Geographical distribution of A-OIV**

As of 9 April 2013, 11 cases in Shanghai, 8 cases in Jiangsu Province, 3 cases in Zhejiang Province and 2 cases in Anhui are reported.

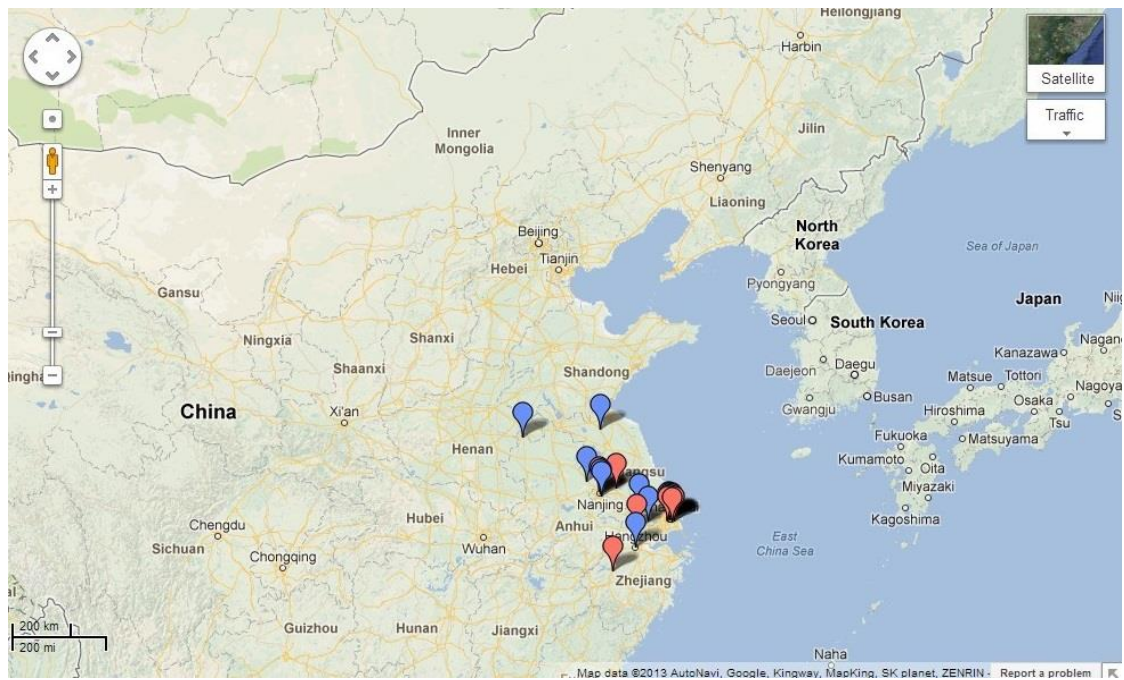

Figure S2: Geographical location of reported infection cases in eastern China (April 9, 2013). Real time updated version can be accessed at *http://goo.gl/maps/ZsVW8*

**Section 2: A-OIV is not directly derived from classical H7N9**

Classical H7N9 viruses have been reported as early as 1999. We performed phylogenetic analysis on HA and NA protein of A-OIV with classical H7N9, and found that A-OIV and classic H7N9 share less similarity compared to other strains such as H7N3 and H7N1. Therefore the A-OIV is not originated from the classical H7N9 by mutation (Fig. S2 and S3).

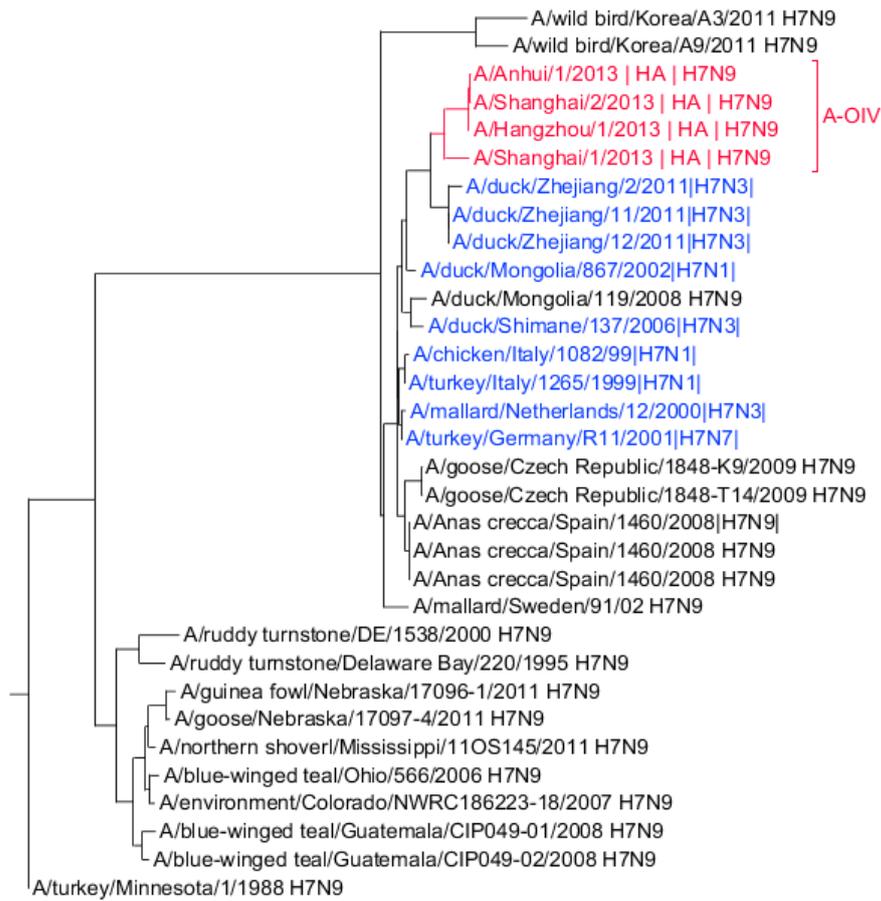

Figure S2: The phylogenetic tree for protein sequences of the HA gene of selected influenza viruses. We selected all A-OIV in human, recently reported H7 family influenza viruses.

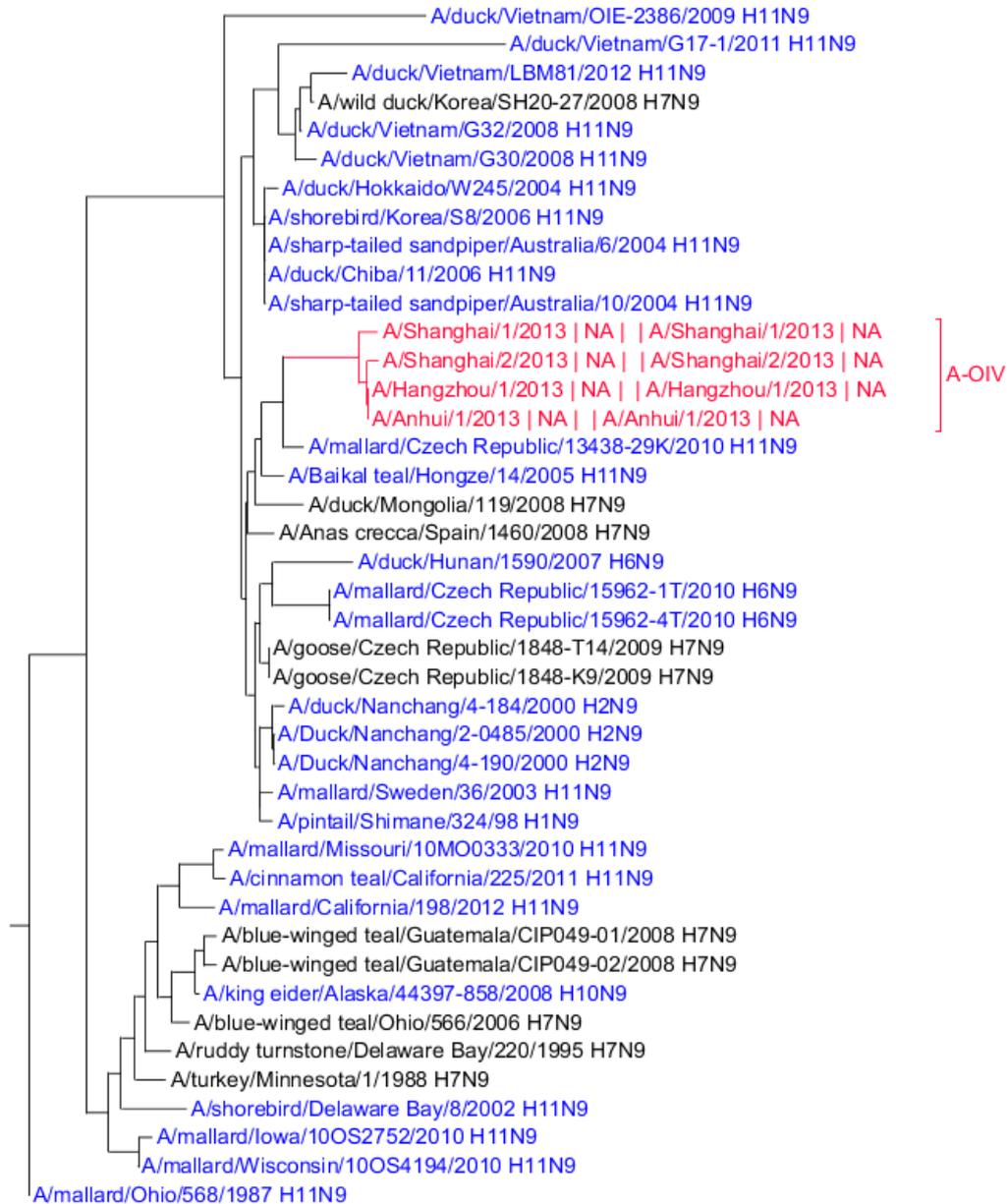

Figure S3: The phylogenetic tree for protein sequences of the NA gene of selected influenza viruses. We selected all A-OIV in human, recently reported N9 family influenza viruses.

**Section 3: Blast analysis**

We blasted every segment of A-OIV against all influenza genomes in the NCBI Influenza Virus Resource. We found that A/duck/Zhejiang/12/2011 strain has the highest score of HA protein with A-OIV, with protein sequence identity >97% (Table 1), which indicates that the A-OIV is a H7 subtype. Blast results show that the neuraminidase (NA) gene is from the H11N9 with protein sequence identity 95% and the six internal proteins are from the H9N2 with protein sequence identity >=98%. The blast results are consistent with phylogenetic analyses that the A-OIV stems from a reassortment of H9N2, H11N9 and H7 family influenza viruses.

| Protein | Highest identity of blast | Collected location (Province, Country) | Sequence identity with A-OIV | Segment coverage | Subtype |
|---------|---------------------------|----------------------------------------|------------------------------|------------------|---------|
| HA | A/duck/Zhejiang/12/2011 | Zhejiang, China | 98% | 100% | H7N3 |
| NA | A/mallard/Czech Republic/13438-29K/2010 | Czech Republic | 95% | 100% | H11N9 |
| PA | A/chicken/Zhejiang/607/2011 | Zhejiang, China | 99% | 100% | H9N2 |
| M1 | A/chicken/Zhejiang/607/2010 | Zhejiang, China | 100% | 100% | H9N2 |
| M2 | A/chicken/Jiangsu/Q3/2010 | Jiangsu, China | 98% | 100% | H9N2 |
| NS1 | A/chicken/Shuanggou/1/2011 | Jiangsu, China | 99% | 100% | H9N2 |
| NS2 | A/chicken/Shuanggou/1/2011 | Jiangsu, China | 100% | 100% | H9N2 |
| PB1 | A/chicken/jiangsu/Q3/2010 | Jiangsu, China | 99% | 100% | H9N2 |
| PB1-F2 | A/chicken/Zhejiang/HJ/2007 | Zhejiang, China | 98% | 100% | H9N2 |
| PB2 | A/chicken/Anhui/HF/2010 | Anhui, China | 99% | 100% | H9N2 |
| NP | A/chicken/Zhejiang/Q1D4/2010 | Zhejiang, China | 99% | 100% | H9N2 |

Table 1: Protein blast results of gene segments of A-OIV strain A/Anhui/1/2013.

**Section 4: Comprehensive phylogenetic analysis of 1193 genomes**

The genome sequences of newly sequenced 7 strains of 2013 A-OIV were achieved in GISAID. 4479 complete avian influenza genomes of any subtype and sampling year was downloaded from the NCBI Influenza Virus Resource on April 6, 2013. The annotation information of each sequence including the subtype, host, year of isolation, geographical location was also downloaded. The protein sequences are used for the following analysis. To remove sequence redundancy, we listed all isolates in which the protein sequence of in segment 1 (PB2) was at least on amino acid different from others. Next, to represent the spectrum of influenza A diversity, we sampled the sequence sets. We selected one genome per country, per year, per subtype. The final genome set included in our study contains 1193 complete genome sequences. The protein sequence of PB1, PB2, PA, NP, NA, MP, NS, HA were aligned using MAFFT with default parameters. Phylogenetic trees were inferred using the UPGMA method, plotted in TreeGraph software.

Fig. S4 is the phylogenetic tree of HA proteins. We found that A-OIV is in the H7 family,

therefore, we conclude that the HA protein of A-OIVs is H7 subtype. Fig. S5 is the phylogenetic tree of NA proteins. We found that A-OIV is clustered within the N9 family, therefore, we conclude that the HA protein of A-OIV is N9 subtype. The two most related strains are H11N9 and H7N9, represented by A/mallard/Czech Republic/13438-29K/2010(H11N9) and A/Wild Bird/Korea/A14/2011(H7N9). Further, we made a phylogenetic tree of only N9 subtype of Fig. S5, as shown in Fig. S15, we found that A-OIV is clustered together with H11N9 lineage; However, we need more evidence to determine the origin of NA gene segment of A-OIV. Figs. S6-S14 are the phylogenetic trees of internal proteins (M1, PB2, PB1-F2, PB1, PA, NS2, NS1, NP, M2). These 9 phylogenetic trees show a consistent pattern that A-OIV is clustered together with H9N2, therefore, we suggest that the six internal genes of A-OIV are originated from H9N2. These results support the Figure 1 in main text.

We gratefully acknowledge the authors, originating and submitting laboratories of the sequences from GISAID's EpiFlu™ Database on which this research is based. The list is detailed below.

| Isolate ID | Country | Collection date | Isolate name | Originating Laboratory | Submitting Laboratory | Authors |
|---|---|---|---|---|---|---|
| EPI_ISL_138977 | China | 2013-03-24 | A/Hangzhou/1/2013 | Hangzhou Center for Disease Control and Prevention | Hangzhou Center for Disease Control and Prevention | Li,J; Pan,JC; Pu,XY; Yu,XF; Kou,Y; Zhou,YY |
| EPI_ISL_138739 | China | 2013 | A/Anhui/1/2013 | WHO Chinese National Influenza Center | WHO Chinese National Influenza Center | |
| EPI_ISL_138738 | China | 2013 | A/Shanghai/2/2013 | WHO Chinese National Influenza Center | WHO Chinese National Influenza Center | |
| EPI_ISL_138737 | China | 2013 | A/Shanghai/1/2013 | WHO Chinese National Influenza | WHO Chinese National Influenza | |

| | | | | Center | Center | |
|---|---|---|---|---|---|---|
| EPI_ISL_138985 | China | 2013-04-02 | A/Pigeon/Shanghai/S1069/2013 | Harbin Veterinary Research Institute | Harbin Veterinary Research Institute | Kong, Huihui |
| EPI_ISL_138983 | China | 2013-04-03 | A/Chicken/Shanghai/S1053/2013 | Harbin Veterinary Research Institute | Harbin Veterinary Research Institute | Kong, Huihui |